\documentclass[fleqn,usenatbib]{mnras}

\usepackage{newtxtext,newtxmath}

\usepackage[T1]{fontenc}

\DeclareRobustCommand{\VAN}[3]{#2}
\let\VANthebibliography\thebibliography
\def\thebibliography{\DeclareRobustCommand{\VAN}[3]{##3}\VANthebibliography}

\usepackage{graphicx}	
\usepackage{amsmath}	
\usepackage{caption}
\usepackage{xcolor}
\usepackage{subcaption}
\usepackage[normalem]{ulem} 

\defcitealias{Donkelaar:2022ab}{Paper~I}

\newcommand{\soutPC}{\bgroup\markoverwith{\textcolor{cyan}{\rule[0.5ex]{2pt}{1pt}}}\ULon}

\newcommand{\soutdif}{\bgroup\markoverwith{\textcolor{orange}{\rule[0.5ex]{2pt}{1pt}}}\ULon}
\newcommand{\soutnew}{\bgroup\markoverwith{\textcolor{red}{\rule[0.5ex]{2pt}{1pt}}}\ULon}

\title[Hybrid nuclear star cluster formation]{Stellar cluster formation in a Milky Way-sized galaxy at $z>4$ -- II. A hybrid formation scenario for the nuclear star cluster and its connection to the nuclear stellar ring}

\author[F. van Donkelaar et al.] 
{Floor van Donkelaar,$^{1}$\thanks{floor.vandonkelaar@uzh.ch} Lucio Mayer,$^{1}$ Pedro R. Capelo,$^{1}$ Tomas Tamfal,$^{1}$ Thomas R. Quinn$^{2}$ \newauthor and Piero Madau$^{3}$\\
$^1$Department of Astrophysics, University of Zurich, Winterthurerstrasse 190, CH-8057 Z\"urich, Switzerland\\
$^2$Astronomy Department, University of Washington, Seattle, WA 98195, USA\\
$^3$Department of Astronomy and Astrophysics, University of California, 1156 High Street, Santa Cruz, CA 95064, USA}

\date{Accepted XXX. Received YYY; in original form ZZZ}

\pubyear{2024}

\begin{document}
\label{firstpage}
\pagerange{\pageref{firstpage}--\pageref{lastpage}}
\maketitle

\begin{abstract}
Nuclear star clusters (NSCs) are massive star clusters found in the innermost region of most galaxies. While recent studies suggest that low-mass NSCs in dwarf galaxies form largely out of the merger of globular clusters and NSCs in massive galaxies accumulate mass primarily through central star formation, the formation channel  of the Milky Way's NSC is still uncertain. In this work, we use GigaEris, a high resolution $N$-body, hydrodynamical, cosmological ``zoom-in'' simulation, to investigate a possible formation path of the NSC in the progenitor of a Milky Way-sized galaxy, as well as its relation to the assembly and evolution of the galactic nuclear region. We study the possibility that bound, young, gas-rich, stellar clusters within a radius of 1.5~kpc of the main galaxy's centre at $z>4$ are the predecessors of the old, metal-poor stellar population of the Milky Way's NSC. We identify 47 systems which satisfy our criteria, with a total stellar mass of  $10^{7.5}$~M$_{\sun}$. We demonstrate that both stellar cluster accretion and in-situ star formation will contribute to the formation of the NSC, providing evidence for a hybrid formation scenario for the first time in an $N$-body, hydrodynamical, cosmological ``zoom-in'' simulation. Additionally, we find that the gas required for in-situ star formation can originate from two pathways: gas-rich stellar clusters and gas influx driven by large-scale non-axisymmetric structures within the galaxy. This is partly supported by the presence of a stellar ring, resulting from gas dynamics,  with properties similar to those of the Milky Way’s nuclear stellar disc.
\end{abstract}

\begin{keywords}
galaxies: formation -- galaxies: high-redshift -- galaxies: nuclei -- methods: numerical
\end{keywords}


\section{Introduction}\label{sec:intro}

The Galactic Centre of the Milky Way (MW) is an excellent laboratory for studying phenomena and physical processes that may be occurring in many other galactic nuclei.  The MW's nuclear star cluster (NSC) and nuclear stellar disc (NSD) are the main features of the Galactic Centre. Nevertheless, their observation is hampered by the extreme source crowding and high extinction. Hence, their relation and formation scenario are not fully clear yet \citep[][]{Schodel:2021aa, Nogueras:2021aa}.

NSCs are extremely dense and massive star clusters occupying the innermost region of a majority of galaxies of all types \citep[e.g.][]{Carollo:1997aa, Matthews:1999aa, Boker:2002aa, Cote:2006aa}. They are more luminous than globular clusters \citep[GCs; e.g.][]{Boker:2010aa} and have masses of the order of $\sim$$10^4$--$10^9$~M$_{\sun}$ \citep[][]{Walcher:2005aa, Fahrion:2020aa, Fahrion:2021aa} and effective radii of the order of 1--20~pc \citep[see][and references therein]{Neumayer:2020aa}. Many NSCs appear to be non-spherical. This is supported by observations of edge-on spirals which identified elongated, i.e. disc-like, structures in NSCs that are well-aligned with the disc of their host galaxies \citep{Seth:2006aa}. The NSCs with a higher stellar mass tend to be more flattened than lower-mass NSCs \citep[e.g.][]{Spengler:2017aa, Georgiev:2014aa}.

NSCs exist in very different host environments \citep[][]{Neumayer:2020aa}, which raises the question of whether NSC formation is controlled by similar processes in all galaxy types, or if NSCs follow evolutionary paths that depend on the properties of their host galaxy. As the NSC is one the main features of the MW's Galactic Centre, its formation path is of exceptional interest for the understanding of physical processes that occur in the central region of the MW. The MW's NSC extends up to hundreds of arcseconds across from the central supermassive black hole  and is believed to have a stellar mass of a few $10^7$~M$_{\sun}$ \citep[e.g.][]{Launhardt:2002aa, Feldmeier:2014aa, schodel:2014ab, Chatzopoulos:2015aa, Fritz:2016aa, Fledmeier:2017aa}.

The formation and evolution path of the MW's NSC  are still unknown. There are two main hypotheses that have been suggested for the formation of the cluster: (a) through GC accretion; (b) through in-situ star formation \citep[SF;][]{Neumayer:2020aa}. In the GC-accretion scenario, the NSC forms out of the gas-free merger of GCs that spiral into a galaxy’s centre due to dynamical friction \citep[DF; e.g.][see also \citealt{Clarke:2019aa} for clump accretion]{Tremaine:1975aa, Capuzzo:1993aa, Capuzzo:2008aa, Agarwal:2011aa, Arca:2014aa, Gnedin:2014aa}. An NSC formed through GC accretion is expected to reflect properties typical of GCs, which are in general characterised by simple SF histories (SFHs), low metallicity, and a high fraction of old stars. The alternative formation path, in-situ SF, considers an NSC to form directly at the galactic centre out of star-forming gas \citep[e.g.][]{Milsavljevi:2004aa, McLaugh:2006aa, Bekki:2007aa}. In this nuclear SF scenario, gas falls into the nucleus and then transforms into stars \citep{Loose:1982}. As SF can proceed in several episodes during a galaxy’s evolution, an NSC formed in this way will exhibit a more complex SFH and is expected to have a low mass fraction of old and metal-poor stellar populations in comparison to NSCs that formed through GC accretion \citep[e.g.][]{Antonini:2012aa, Feldmeier:2015aa, Arca:2020aa, Fahrion:2020aa, Fahrion:2021aa, Fahrion:2022aa}.

Both processes may contribute with different weights in different galaxies, meaning that both mechanisms could contribute to the build-up of NSCs over a Hubble time \citep[see, e.g.][]{Guillard:2016aa}, through a hybrid formation scenario. The infall of GCs alone is not a viable formation scenario for the NSCs in massive early-type galaxies with stellar mass $M_{\star} > 10^9$~M$_{\sun}$, as the observed metallicity is too high and the SFH shows an ongoing process \citep[][]{Walcher:2005aa, Fahrior:2019aa, Pinna:2021aa}. However, mergers and accretion of gas-rich young stellar clusters \citep[][]{Figer:2002aa, Paudel:2020aa} could still provide a viable formation mechanism. Such a formation pathway can lead to similar stellar population properties as those observed in the MW's NSC.

In spite of its proximity, the observation of the MW's NSC is restrained by the extreme source crowding and the high interstellar extinction that limits its analysis to the infrared regime \citep[e.g.][]{Nishiyama:2008aa, schodel:2010aa, Nogueras:2018aa, Nogueras:2020ab}. Consequently, the relation between the NSC and the NSD is also not well understood yet \citep[see, e.g.][]{Launhardt:2002aa, Schodel:2020aa, Nogueras:2020aa}. There is some evidence that the NSC and the NSD may host different stellar populations with different SFHs \citep[e.g.][]{Nogueras:2020aa, Schodel:2020aa, Schultheis:2021aa}. Both components seem to have a predominantly old stellar population, with the initial starburst followed by several billion years of quiescence. Nevertheless, there is some evidence for a $\sim$3~Gyr old intermediate-age population in the NSC, which cannot be found in the NSD. On the other hand, there is evidence for a $\sim$1~Gyr old SF event associated to the NSD, that is not found when analysing the stellar population of the NSC \citep[see][]{Schodel:2020aa, Nogueras:2020aa, Nogueras:2021aa}. Because of the differences, understanding the formation paths of both systems can help us interpret the physical processes that occur in galactic nuclei.

This was initiated by \citet{Becklin:1982aa}, who showed that the region within 2~pc of the MW's Galactic Centre is largely devoid of interstellar matter and is surrounded by a dust ring or disc. More recent observations and simulations have determined this to be an NSD which includes stars and is star-forming \citep[e.g.][]{Launhardt:2002aa, Sungsoo:2012aa, Schultheis:2021aa}. The NSD of the MW extends up to a radius of 220~pc with a scale height of $\sim$50~pc \citep[][]{Piere:2000aa, Launhardt:2002aa,Nogueras:2020aa, Gallego:2020aa}. NSDs are also detected in extragalactic systems and are quite common in early-type galaxies \citep[][]{Pizella:2002aa, Gadotti:2019aa}. However, given that no NSDs have been clearly identified in late-type galaxies and provided that the MW is a barred spiral galaxy, it is plausible that the NSD of the MW is rather a mixture of a nuclear star-forming ring (the nuclear stellar ring; NSR) and a nuclear spiral \citep[][]{Schodel:2021aa}. Nevertheless, the fact that the MW has not had any major merger in the past $\sim$10 Gyr \citep[e.g][]{Wyse:2001aa,Helmi:2018aa, Renaud:2021aa, Sotillo:2022aa}  makes the existence of a nuclear disc, such as those observed in S0 galaxies, still a possibility in the MW \citep[see, e.g.][]{Sarzi_et_al_2015,Galan-deAnta_et_al_2023a,Galan-deAnta_et_al_2023b}. The bulk of stars in the MW's NSD is old and formed at least 8~Gyr ago, followed first by a phase of quiescence and then by recent SF activity \citep[about 1~Gyr ago, when 5 per cent of the mass of the NSD was formed very quickly;][]{Nogueras:2019aa, Nogueras:2020aa, Nogueras:2021aa, Nogueras:2023aa, Schodel:2020aa}.

Galactic bars can lead to the creation of substructures in the nuclear region of disc galaxies such as NSDs and NSRs by redistributing angular momentum \citep[][]{Combes:1985aa}. There is now growing evidence that these substructures in the galactic centre are built from gas that was funnelled to the centre by the bar \citep[e.g.][]{Contopoulos:1989aa, Binney:1991aa,Knapen:1999aa, kim:2011aa, Fragkoudi:2016aa}. Another formation scenario of NSDs could be galaxy mergers \citep[e.g.][]{Mayer:2008aa}; these, however, fail to reproduce the sizes of typically observed NSDs in nearby galaxies \citep[][]{Schultheis:2021aa}.

Moreover, since we do not detect an age-metallicity relation in the solar neighbourhood, clusters of stars are expected to undergo radial migration \citep[][]{Sellwood:2002aa, Haywood:2008aa, Roskar:2008aa, Schonrich:2009aa}. Therefore, one could expect that gas-rich stellar clusters migrated from the outer parts of the Galaxy towards the Galactic Centre. For example, in \citet{Schonrich:2009aa}, the stars are trapped on to a resonant co-rotation with spiral arms and may migrate inwards and outwards along the spiral waves. The thin disc is expected to start forming the earliest assembly stage of a galaxy \citep[e.g.][see also \citealt{Agertz2021, Silva:2021aa, Michael:2022aa, vanDonkelaar:2022aa} for the early formation of a thin-disc component through co-formation of the discs]{Tamfal:2022aa}, therefore a significant population of old thin-disc stars could have influenced the formation of the NSC and NSD.

From a theoretical point of view, hydrodynamic simulations of MW analogs have shown that the formation of the bar can trigger gas funnelling to the centre of the MW, forming a kinematically cold, rotating NSD \citep[e.g.][]{Fux:1999aa, Li:2015aa, Ridley:2017aa, Sormani:2019aa, Tress:2020aa, Moon:2021aa, Sormani:2022aa}. Additionally, a large number of simulations and semi-analytic models have calculated the efficiency of DF for a range of starting conditions for GC systems and their host galaxies, confirming that DF provides a plausible mechanism to form an NSC \citep[e.g.][]{Capuzzo:1993aa, Lotz:2001aa}. Detailed $N$-body simulations of the infall of GCs through DF have shown consistency with the flattening and the kinematic properties of observed NSCs, but that matching the kinematics and luminosity function of the stellar clusters likely requires roughly half of the NSC mass to come from in-situ SF and accreted gas \citep[e.g.][]{Hartmann:2011aa, Antonini:2012aa, Tsatsi:2017aa}. Furthermore, semi-analytic models that follow the evolution of GC systems with DF to track the growth of the NSCs have been able to reproduce some of the properties, like mass and radius, of both the present-day NSCs and the GC systems \citep[e.g.][]{Antonini:2013aa, Gnedin:2014aa}. \citet{bekki:2010aa} carried out $N$-body simulations of the orbital decay of stellar clusters in the background of field stars in a disc galaxy embedded in a dark matter (DM) halo, finding that NSCs could have formed from stars delivered by inspiralling stellar clusters.

In this work, we aim to investigate the possibility of the component of old, metal-poor stars within the MW's NSC being formed through the infall of gas-rich stellar clusters with an $N$-body, hydrodynamical, cosmological ``zoom-in''  simulation of unprecedented resolution, GigaEris \citep[][]{Tamfal:2022aa}. Additionally, we explore the relation between the MW's NSC and its NSD, and how their formation may have been connected. For that reason, we delve into the properties of young, gas-rich, stellar clusters at $z>4$ in the nuclear region of an MW-sized galaxy. The layout of this paper is the following: Section~\ref{sec:method} briefly summarises the simulation setup and describes how we identified the clusters. In Section~\ref{sec:results}, we present the simulation results, first with a focus on the properties of the possible NSC predecessors (NSCPs) at $z= 4.4$ and the possibility that these stellar clusters will form an NSC with the properties of the MW's NSC, and then presenting the NSR within the simulation and its link to the NSC. Finally, we discuss our results in Section~\ref{sec:disc} and conclude in Section~\ref{sec:conc}.

\section{Methods}\label{sec:method}

\subsection{Simulation code and initial conditions}

We base our analysis on an $N$-body, hydrodynamical, cosmological ``zoom-in'' simulation of a MW-like galaxy, GigaEris \citep[][]{Tamfal:2022aa}, carried out with the $N$-body smoothed-particle hydrodynamics (SPH) code \textsc{ChaNGa} \citep[][]{Jetley:2008aa,Jetley:2010aa, Menon:2015aa}. A brief summary of the numerical recipes is provided below; one can find a more detailed discussion on the set-up in \citet{Tamfal:2022aa}.

The GigaEris simulation follows a Galactic-scale halo identified in a low-resolution, DM-only simulation at $z = 0$ in a periodic cube of side 90 cMpc. It was chosen to have a similar mass as that of the MW and a rather quiet late merging history. This method is similar to the way the galaxy halo was selected in the original Eris suite in \citet{Guedes:2011aa}. Next, the selected halo was re-simulated at several orders of magnitude higher resolution than the DM-only simulation, adding gas particles as well as the necessary short-wavelength modes. The initial conditions were generated with the \textsc{MUSIC} code \citep[][]{Hahn:2011aa}, with 14 levels of refinement and the cosmological parameters $\Omega_{\rm m}$ = 0.3089, $\Omega_{\rm b}$ = 0.0486, $\Omega_{\Lambda}$ = 0.6911, $\sigma_8$ = 0.8159, $n_{\rm s}$ = 0.9667, and $H_0$ = 67.74~km~s$^{-1}$ Mpc$^{-1}$ \citep[see][]{Planck:2016aa}. The gravitational softening of all particles is set to a constant in physical coordinates ($\epsilon_{\rm c} = 0.043$~kpc) for redshifts smaller than $z = 10$ and otherwise evolves as $\epsilon = 11\epsilon_{\rm c}/(1+z)$. The smoothing length is set so that it cannot be smaller than 5 per cent of the gravitational softening. For the final snapshot at $z=4.4$, the DM, gas, and stellar particle numbers in the entire simulation box are $n_{\rm DM} = 5.7 \times 10^8$, $n_{\rm gas} = 5.2 \times 10^8$ (with a mean gas mass of $m_{\rm gas} = 1099$~M$_{\sun}$), and $n_{\star} = 4.4 \times 10^7$ (with a mean stellar mass of $m_{\star} = 798$~M$_{\sun}$), respectively. 

Each star particle is created stochastically with an initial mass of $m_{\star} = 1026$~M$_{\sun}$ using a simple gas density and temperature threshold criterion \citep[][]{Stinson:2006aa}, with $n_{\rm SF} > 100$~$m_{\rm H}$~cm$^{\text{-} 3}$ and $T_{\rm SF} < 3 \times 10^4$~K,  and the gas particle that spawns the new star has its own mass reduced accordingly. A star particle represents an entire stellar population with its own \citet{Kroupa:2001aa} initial stellar mass function (IMF). The SF  proceeds at a rate which is given by

\begin{equation}
    \frac{\rm{d} \rho_{\star}}{\rm{d}t} = \epsilon_{\rm SF} \frac{\rho_{\rm gas}}{t_{\rm dyn}},
\end{equation}

\noindent with $\rho_{\star}$ indicating the stellar density, $\rho_{\rm gas}$ the gas density, $t_{\rm dyn}$ the local dynamical time, and $\epsilon_{\rm SF}$ the SF efficiency, which is set to $0.1$. The code solves for the non-equilibrium abundances and cooling of H and He species \citep[assuming self-shielding and a redshift-dependent radiation background;][]{Pontzen:2008aa,Haardt:2012aa}, whereas the cooling from the fine structure lines of metals is calculated in photoionization equilibrium from the same radiation background (assuming no self-shielding; see \citealt{Capelo:2018aa} for a discussion), using tabulated rates from Cloudy \citep{Ferland:2010aa, Ferland:2013aa} and following the method described in \citet{Shen:2010aa, Shen:2013aa}.

Feedback from supernovae (SNae) Type Ia  is implemented by injecting energy and a fixed amount of mass and metals, independent of the progenitor mass, into the surroundings \citep[see][]{Thielemann:1986, Stinson:2006aa}, whereas SNae Type II (SNII) feedback is implemented following the delayed-cooling recipe of \citet{Stinson:2006aa}, with metals and energy, $\epsilon_{\rm SF} =10^{51}$~erg, being injected per event into the interstellar medium  as thermal energy, according to the ‘blastwave model’ of \citet{Stinson:2006aa}. For each SNII event, a given amount of oxygen and iron mass, dependent on the mass of the star, is injected into the surrounding gas \citep{Woosley:1995aa, raiteri:1996aa}. The stars with masses between 8 and 40~M$_{\sun}$ will explode as SNII, whereas stars with masses between 1 and 8~M$_{\sun}$ do not explode as SNae but release part of their mass as stellar winds, with the returned gas having the same metallicity of the low-mass stars.

\subsection{Cluster finding}

We use the adaptive mesh \textsc{AMIGA Halo Finder} \citep[\textsc{AHF};][]{Gill:2004aa, Knollmann:2009aa} to identify the gas-rich stellar substructures that possibly could be connected to the formation of the NSCs in the final simulation step  at $z=4.4$. The clusters were selected in an identical way as in  \citeauthor{Donkelaar:2022ab} (\citeyear{Donkelaar:2022ab}; hereafter \citetalias{Donkelaar:2022ab}), namely with a minimum threshold of 64 baryonic particles in the virial radius per cluster and 0 subclusters inside the identified cluster. Furthermore, to validate that the clusters are bound, we have calculated the binding energies of the identified clusters and removed unbound clusters from the set. Hereafter the word cluster in this paper refers to a bound object found by \textsc{AHF}.

\section{Results}\label{sec:results}

\subsection{Classifying stellar systems}\label{sec:class}

\begin{figure}
\centering
\setlength\tabcolsep{2pt}
\includegraphics[ trim={0cm 0cm 0cm 0cm}, clip, width=0.46\textwidth, keepaspectratio]{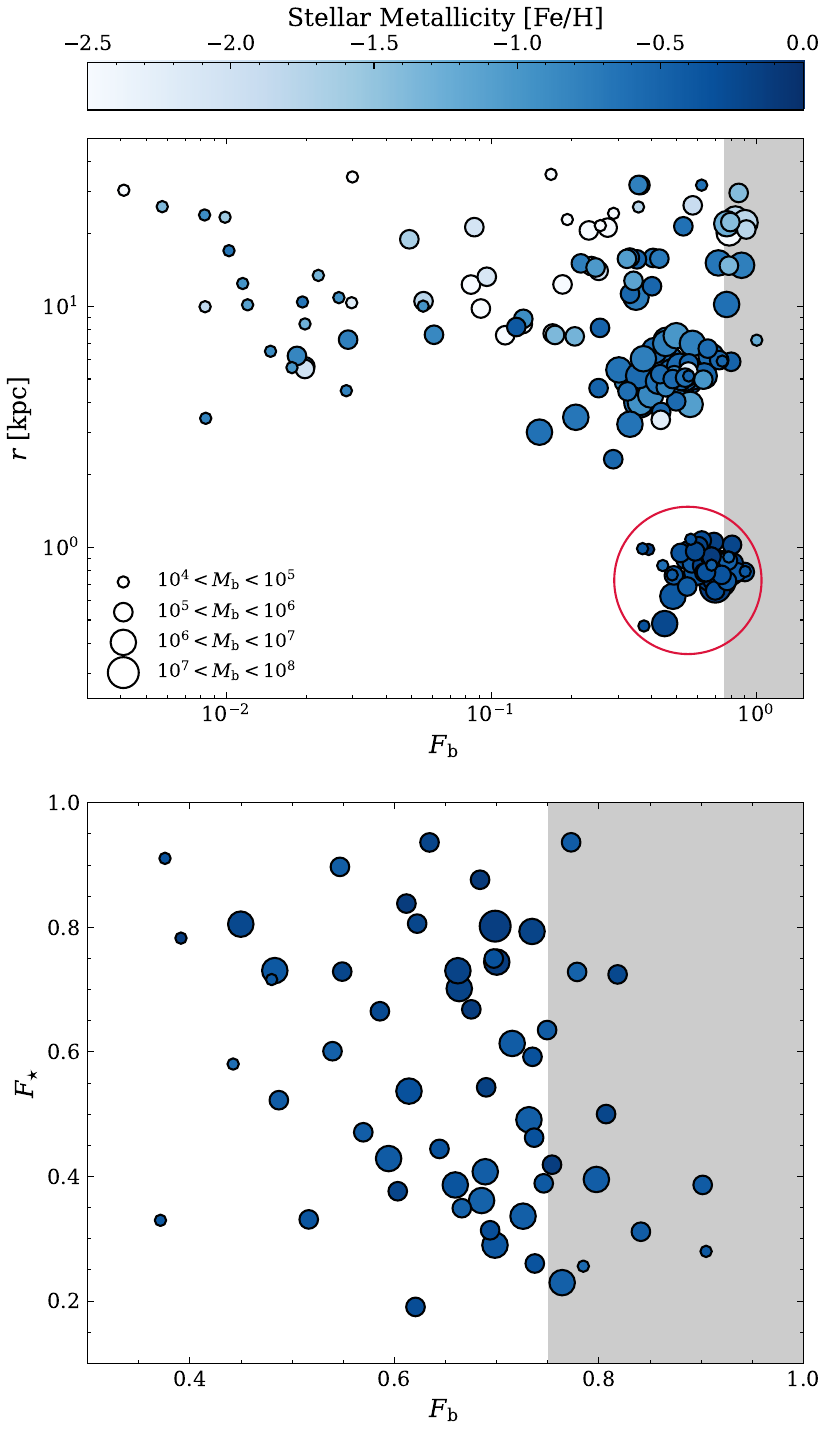}
\caption{Top panel: the object's distance away from the galactic centre of the main galaxy halo plotted against its baryonic fraction, $F_{\rm b}$, for all 213 halos identified with a baryonic mass range $10^4$--$10^8$~M$_{\sun}$ at $z=4.4$. The sizes of the markers indicate the mass range of the clusters. Bottom panel: the stellar fraction, $F_{\star}$, of the 56 clusters within the red circle of the top panel is plotted against the baryonic fraction. The gray-shaded area indicates the region where $F_{\rm b} \geq 0.75$ (one of the criteria used to categorize a cluster as a possible proto-GC; see \citetalias{Donkelaar:2022ab}). The colour bar represents the stellar metallicity of the clusters.}
\label{fig:F_z}
\end{figure}

To investigate the possibility that gas-rich stellar clusters at $z>4$ are NSCP candidates within our simulation, we extract all \textsc{AHF} identified substructures in the simulated box with a baryonic mass range $10^4$--$10^8$~M$_{\sun}$, giving us a total of 213 halos. This criterion helps us exclude clusters with fewer than 10 baryonic particles and eliminates those clusters that already surpass the mass of the MW's NSC at $z=0$. For each of the substructures, the baryonic mass fraction,

\begin{equation}\label{eq:fb}
    F_{\rm b} = \frac{M_{\star}+M_{\rm gas}}{ M_{\rm cluster}},
\end{equation}

\noindent and stellar mass fraction,

\begin{equation}\label{eq:fs}
    F_{\star} = \frac{M_{\star}}{M_{\star}+M_{\rm gas}},
\end{equation}

\noindent are calculated, where $M_{\rm cluster}$ is the sum of the baryonic and DM masses within the clusters, $M_{\star}$ the stellar mass, and $M_{\rm gas}$ the gas mass, all computed at half the virial radius. This approach is similar to that of \citetalias{Donkelaar:2022ab}. The distance away from the galactic centre of the main galaxy of the identified clusters is plotted against their baryonic mass fraction in the top panel of Figure~\ref{fig:F_z}. In \citetalias{Donkelaar:2022ab}, we selected all clusters with an $F_{\rm b} \geq 0.75$ as proto-GC systems (excluding systems with a $\sigma_{\star} < 20$~km~s$^{-1}$): this region has been indicated by the gray-shaded area in the Figure. In the top panel, we see a new group of interesting clusters, all being very close to the galactic centre of the main galaxy ($r <1.5$~kpc), with a relatively high baryonic fraction ($F_{\rm b}>0.35$), and with a similar stellar metallicity (around solar values).\footnote{In this work, we compute the abundance ratios (e.g. [Fe/H] and [O/Fe]) normalising them to the solar values provided by \citet{Asplund:2009aa}.} We select all bound clusters within this region, indicated by the red circle in the top panel of the Figure, as possible NSCPs.

In the bottom panel of Figure~\ref{fig:F_z}, the stellar mass fraction is plotted against the baryonic mass fraction for these 56 possible NSCPs. From this Figure, we can conclude that, even though this group of stellar clusters are in a similar region of the main galaxy and have a similar metallicity, they have a wide variety of stellar and baryonic mass fractions. Moreover, we find that all possible NSCPs still include gas and that approximately half of the clusters are dominated by their gas mass.

We have not used a minimum baryonic mass fraction as part of our selection criteria. This is because the NSCPs are not the final product. Their baryonic mass fraction can change during their evolution, especially when they fall into the high-density central region of the galaxy. This is for example similar to what happened to ``The Imposter'' in \citetalias{Donkelaar:2022ab}, which lost all of its  DM while spiralling into the MW analog.

\subsubsection{Dynamical friction time-scale}\label{sec:dynamicaltime}

\begin{figure}
\centering
\setlength\tabcolsep{2pt}%
\includegraphics[ trim={0cm 0cm 0cm 0cm}, clip, width=0.48\textwidth, keepaspectratio]{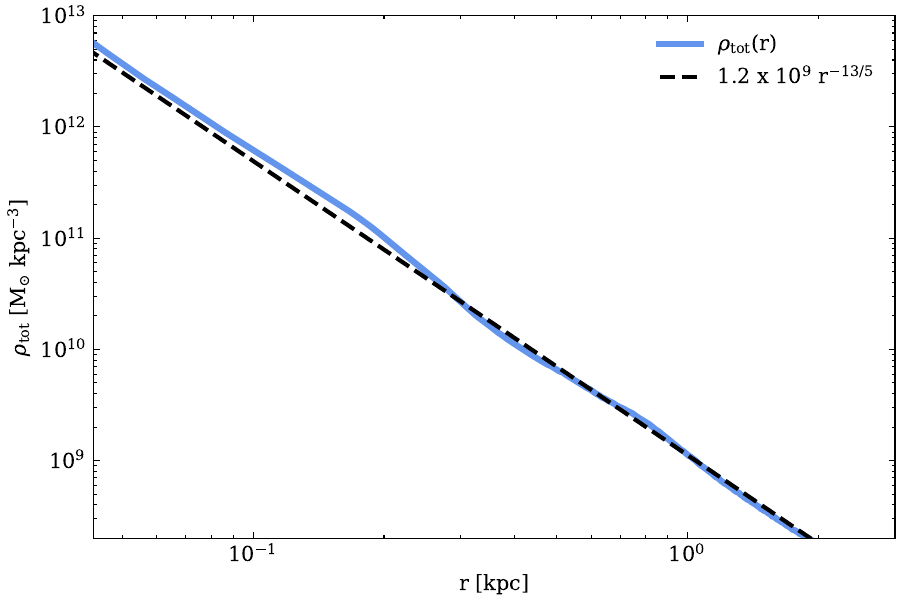}
\caption{Total density profile of the central region of our simulated galaxy at $z = 4.4$ (blue, solid line). The black, dashed line shows the best-fitting profile, $\rho_{\rm tot}(r) = 1.2 \times 10^9 (r/{\rm kpc})^{-13/5}$~M$_{\sun}$~kpc$^{-3}$.}
\label{fig:hern}
\end{figure}

In an attempt to estimate the time taken by a given possible NSCP to decay to the centre of the galaxy, we first approximate the inner parts of the total galactic density profile with an analytical model. As shown in Figure~\ref{fig:hern}, the best fitting model can be described as $\rho_{\rm tot}(r) = \rho_0 (r/r_0)^{-13/5}$, where $\rho_0 = 1.2 \times 10^9$~M$_{\sun}$~kpc$^{-3}$ and $r_0 = 1$~kpc. The solid, blue line shows the central total density profile when assuming a spherically averaged distribution of the galaxy between 0.02 and 2~kpc, whereas the black, dashed line shows the power-law analytical model discussed.

The corresponding time-scale for a cluster of  mass $M_{\rm cluster}$ (computed at half the virial radius) in motion inside such a matter distribution to decay from an initial radial distance, $r_{\rm i}$, from the galactic centre to a final distance, $r_{\rm f}$, can be estimated using the \citet{chandra:1943aa}'s DF formalism, which gives us the force to which the cluster is subjected \citep[see also][]{Binney:2008aa}:

\begin{equation}\label{chandra}
    \mathbf{F}_{\rm DF} = -16 \pi^2 G^2 M^2_{\rm cluster} m_{\rm a} \ln{ \Lambda} \left[ \int_0^v  v_{\rm a}^2 f(v_{\rm a}) {\rm d}v_{\rm a} \right] \frac{\mathbf{v}}{v^3},
\end{equation}

\noindent where $\mathbf{v}$ is the velocity of the cluster relative to the background, $m_{\rm a}$ is the individual mass of the particles in the background and $v_{\rm a}$ is their velocity, $f(v_{\rm a}){\rm d}v_{\rm a}$ is the number of particles with velocity between $v_{\rm a}$, and $v_{\rm a}+{\rm d}v_{\rm a}$. Lastly, $\ln\Lambda$ is the Coulomb logarithm and $G$ is the gravitational constant.

When we consider an isotropic distribution function for the velocities and $v$ to be considerably larger than the velocities of the surrounding environment, we can estimate the integral term in Equation~\eqref{chandra} up to infinity (within a factor $\xi$ of order unity), obtaining the number density of particles, $n = \rho_{\rm tot}/m_{\rm a}$, divided by $4 \pi$. The expression for the force can then be described by

\begin{equation}\label{chandra2}
    \mathbf{F}_{\rm DF} = -4 \pi \rho_{\rm tot}(r) G^2 M^2_{\rm cluster} \ln{ \Lambda} \xi \frac{\mathbf{v}}{v^3}.
\end{equation}

For simplicity, we consider the cluster on a circular orbit, so that $v = v_{\rm c}$, where $v_{\rm c}$ is the circular velocity, defined as $v_{\rm c} =\sqrt{G M_{\rm encl}/ r}$, where $M_{\rm encl}$ is the enclosed mass, which in our case can be written as $M_{\rm encl} = 10 \pi \rho_0 r_0^{13/5} r^{2/5}$. The circular velocity can thus be described by

\begin{equation}\label{vc}
    v_{\rm c} = \sqrt{\frac{10 \pi G \rho_0 r_0^{13/5}}{r^{3/5} }}.
\end{equation}

Because the force is perpendicular to the radial direction of the cluster and produces a torque, Equation~\eqref{vc} can be substituted into Equation~\eqref{chandra2} to obtain an expression for the torque's magnitude:

\begin{equation} 
    r F_{\rm DF} = \frac{2}{5} G M_{\rm cluster}^2 \xi \ln{\Lambda} r^{-1}.
\end{equation}

The torque applied on the cluster corresponds to the time variation of its angular momentum, $\mathbf{L}$, leading to a change in its magnitude equal to

\begin{equation}
    \frac{{\rm d}L}{{\rm d}t} = \frac{{\rm d}}{{\rm d}t} (M_{\rm cluster} r v ) = M_{\rm cluster} \sqrt{10 \pi G \rho_0 r_0^{13/5}} \frac{7 \dot{r}}{10 r^{3/10}}.
\end{equation}

We can then combine the two equations above, since we also have ${\rm d}L/{\rm d}t=  -r F_{\rm DF}$, to obtain an equation for $\dot{r}$ and, integrating, we obtain a rough estimate for the DF time-scale:

\begin{equation}
    t_{\rm DF} =  \frac{35 r_0^{13/10}}{34 M_{\rm cluster} \xi \ln{\Lambda} } \sqrt{\frac{10 \pi \rho_0}{G}} r^{17/10}.
\end{equation}

\noindent We set $\xi = 1$ (as done in \citealt{Lima:2017aa} and \citealt{Tamburello:2017aa} in similar calculations). The Coulomb logarithm,  $\ln{\Lambda}$, can be approximated as \citep[][]{Binney:2008aa}

\begin{equation} 
   \Lambda \approx \frac{b_{\rm max} v^2_{\rm typ}}{G M_{\rm cluster}},
\end{equation}

\noindent where $b_{\rm max}$ is the maximum impact parameter and $v_{\rm typ}$ is the typical velocity in the system. For each cluster (of mass $M_{\rm cluster}$), we take $b_{\rm max} = r_{\rm i}$ and $v_{\rm typ}^2 = G M(<r_{\rm i})/r_{\rm i}$, where $M(<r_{\rm i})$ is the total enclosed mass within $r_{\rm i}$ \citep[][]{Binney:2008aa}, and compute $\Lambda$, thus obtaining the DF time-scale.

\begin{figure}
\centering
\setlength\tabcolsep{2pt}%
\includegraphics[trim={0cm 0cm 0cm 0cm}, clip, width=0.48\textwidth, keepaspectratio]{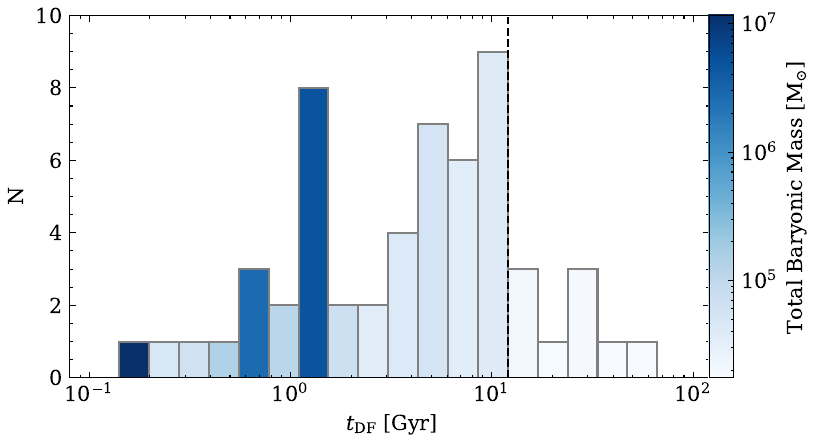}
\caption{The full DF time-scale distribution for all the identified possible NSCPs. The colour bar represents the total baryonic mass within each DF time-scale bin. The dashed vertical line indicates 12~Gyr, which is the time between the last snapshot in the simulation, at $z = 4.4$,  and $z \sim 0$. Nine systems out of 56 have a $t_{\rm DF} > 12$~Gyr.}
\label{fig:tdf}
\end{figure}

The resulting $t_{\rm DF}$ for the clusters is shown in Figure~\ref{fig:tdf}. It shows that 47 (hereafter the NSCPs) of the 56 selected clusters in Figure~\ref{fig:F_z} will decay to the centre within 12~Gyr (indicated by the dashed, vertical line), which is the time between the last snapshot in the simulation, at $z=4.4$, and $z \sim 0$. Moreover, the influence of the disc, bar, or other galactic structures have not been included in this model \citep[see, e.g.][]{Bar:2022aa}. Owing to the galactic bar, for example, a massive object with a similar inclination and distance away from the centre as our clusters can have its inspiral time-scale decreased, as shown by \citet{Bortolas:2020aa,Bortolas:2022aa}. They concluded that for in-plane perturbers there is a clear tendency to decrease the DF time-scale, whereas for perturbers on arbitrary inclinations the effect is stochastic, with both increases and decreases of the DF time-scale similarly possible. Note that our NSCPs are found in a symmetric ring with a  small inclination of $\sim$14$^{\circ}$ with respect to the galactic plane (see Section~\ref{sec:ring}). Therefore, one would expect that the decay time-scale would decrease when adding these structures to the model. We can thus conclude that most of the selected clusters from Figure~\ref{fig:F_z} are consistent with the chosen definition of NSCP. 

\subsection{Properties of the nuclear star cluster predecessors}\label{sec:prop}

\begin{figure}
\centering
\setlength\tabcolsep{2pt}%
\includegraphics[ trim={0cm 0cm 0cm 0cm}, clip, width=0.48\textwidth, keepaspectratio]{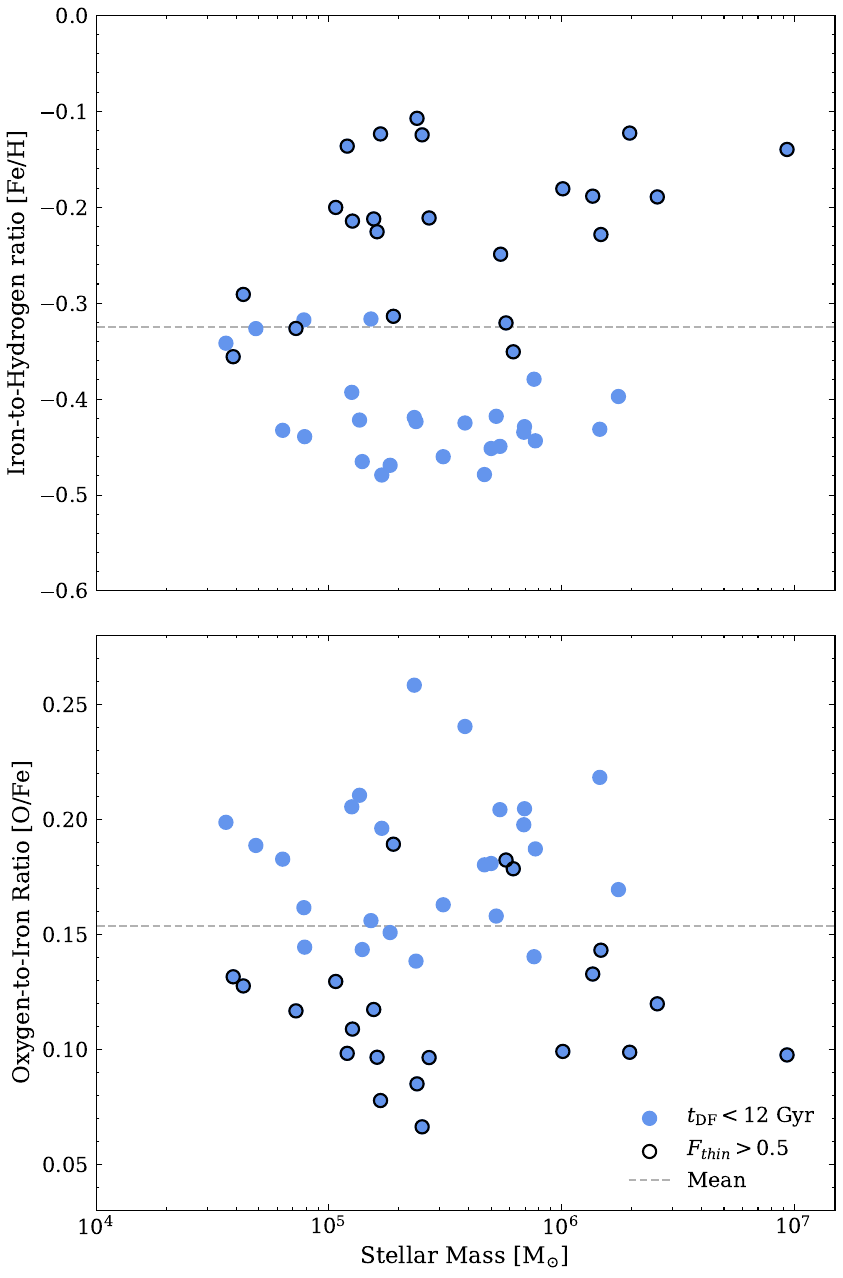}
\caption{Abundance ratios (top panel: [Fe/H]; bottom panel: [O/Fe]) as a function of stellar mass. The blue dots indicate our new stellar cluster sample of 47 systems, defined by applying the selection criterion described in Section~\ref{sec:class} and additionally imposing $t_{\rm DF} < 12$~Gyr (Section~\ref{sec:dynamicaltime}). The dots with a black outline represent the stellar clusters with $F_{\rm thin} > 0.5$ (see Section~\ref{sec:thin}). The gray horizontal lines indicate the mean values.}
\label{fig:prop}
\end{figure}

\begin{figure*}
\centering
\setlength\tabcolsep{2pt}
\includegraphics[ trim={0cm 0cm 0cm 0cm}, clip, width=0.98\textwidth, keepaspectratio]{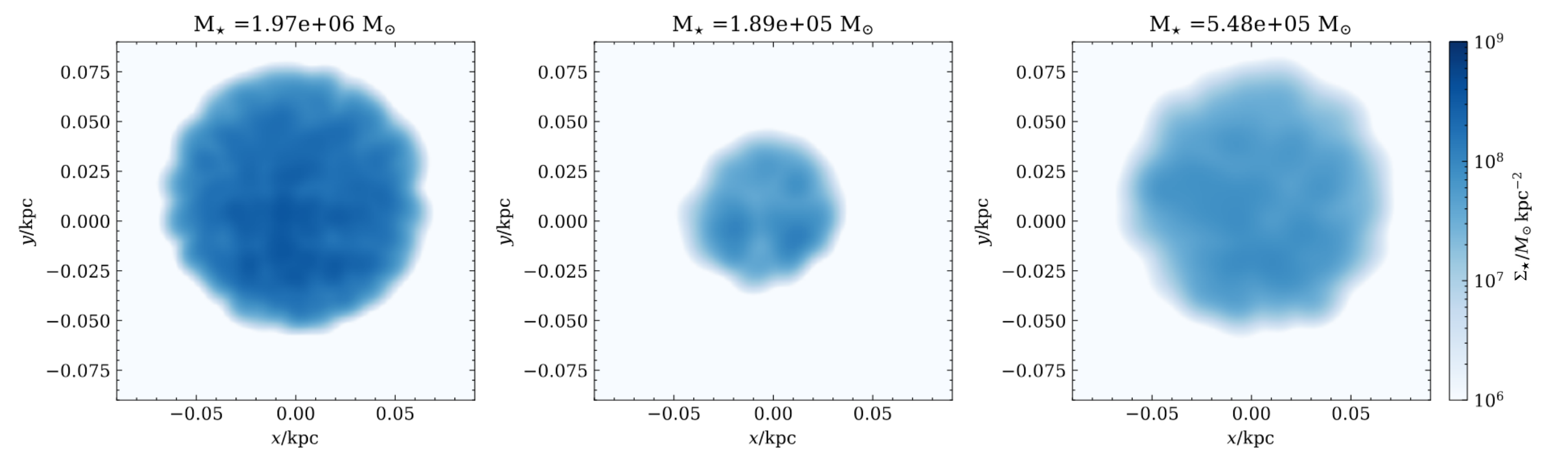}
\caption{Stellar surface density maps of three selected NSCPs at $z=4.4$ centred at their centre of mass. Combining the surface density maps with the knowledge that they follow the radial density profiles of the \citeauthor{King:1972aa} model, we can assume they resemble spherical clumps. The stellar mass of the clusters at this redshift is given by the title of the plot. The gas mass of the shown clusters is between $10^{5.1} \lesssim {M_{\rm gas} [{\rm M}_{\sun}]} \lesssim 10^{5.8}$.}
\label{fig:clumps}
\end{figure*}

We carry on by studying the properties of the selected 47 stellar clusters within 1.5~kpc of the centre of the main galaxy. Together with the DF time-scale estimation, these properties will allow us to investigate the hypothesis that the clusters are indeed predecessors of the NSC. Figure~\ref{fig:prop} shows how some properties of these selected clusters relate to their stellar mass ($M_{\star}$). The total sum of the stellar (baryonic) masses of all 47 NSCPs (we excluded the nine systems with $t_{\rm DF} > 12$~Gyr), is$10^{7.5}$~M$_{\sun}$ ($10^{7.7}$~M$_{\sun}$), which is of the same order of magnitude as the observed stellar mass of the MW's NSC \citep[a few $10^7$~M$_{\sun}$; e.g.][]{Launhardt:2002aa, Feldmeier:2014aa, schodel:2014ab, Chatzopoulos:2015aa, Fritz:2016aa, Fledmeier:2017aa}, which reinforces the possibility of these clusters being NSCPs. Nevertheless, one should note that 28 per cent of this total stellar mass comes from one massive cluster; without this cluster, the combined stellar (baryonic) mass of the NSCPs would be $10^{7.4}$~M$_{\sun}$ ($10^{7.6}$~M$_{\sun}$). This cluster has, discarding the mass, average properties in comparison to the other clusters and is therefore still included as a possible NSCP. Furthermore, all the NSCPs are spherical clumps (as shown by the examples in Figure~\ref{fig:clumps}) and follow the radial density profiles of the \citet{King:1972aa} model.

In the top panel of Figure~\ref{fig:prop}, the stellar metallicity, [Fe/H], is plotted against the stellar mass for the selected sample of NSCPs. The metallicity of the possible NSCPs is spread between $-0.42$ and $-0.11$, with a mean of $-0.32$. The clusters have a lower mean metallicity than what one would expect from today's MW's NSC, which is solar to super solar, and at the lower end of what is expected from galaxies with a stellar mass above $10^9$~M$_{\sun}$ \citep[$-0.5 \lesssim {\rm [Fe/H]} \lesssim 0.5$; e.g.][]{Kacharov:2018aa, Schodel:2020aa}. However, this is reasonable at $z=4.4$, as our clusters are still star-forming at this redshift and thus the mean stellar metallicity of the stars in the NSCPs can still be raised. Furthermore, assuming a hybrid formation scenario for the MW's NSC, stars formed through central SF would also raise the mean metallicity over time. Comparing this to the metallicity of proto-GCs in \citetalias{Donkelaar:2022ab} ($-1.8 \lesssim {\rm [Fe/H]} \lesssim -0.8$), we can assume that the metallicity of these selected stellar clusters is too high to be a proto-GC and closer to what one would expect from the MW's NSC.

\begin{figure}
\centering
\setlength\tabcolsep{2pt}%
\includegraphics[ trim={0cm 0cm 0cm 0cm}, clip, width=0.49\textwidth, keepaspectratio]{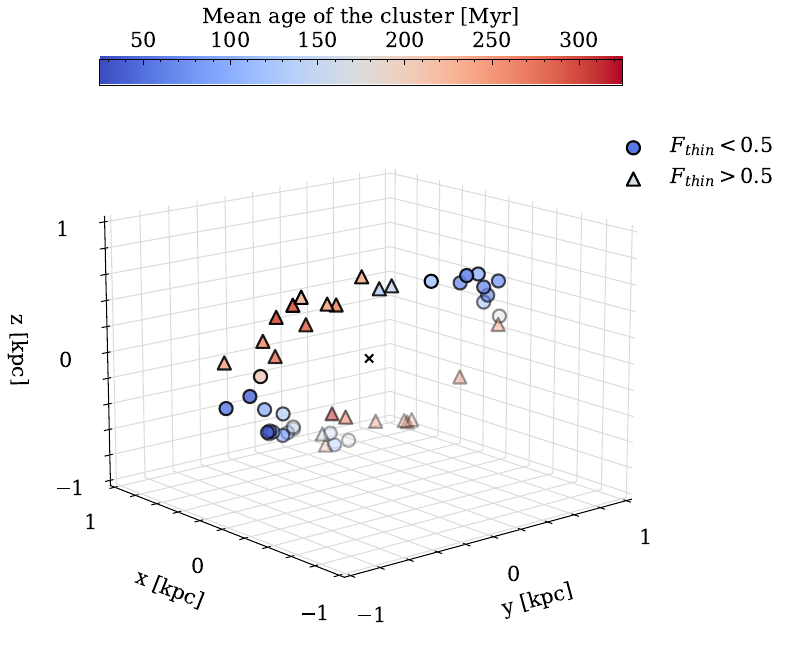}
\caption{The location of the 47 NSCPs in the $x$-$y$-$z$-space, with the stellar galactic angular momentum pointing in the $+z$ direction. The black `$\times$' indicates the centre of the main galaxy halo. The colour bar represents the mean age of the stars within the clusters.}
\label{fig:thindisc}
\end{figure}

The bottom panel of Figure~\ref{fig:prop} shows the [O/Fe] ratio of the possible NSCPs, all within $0.02 \lesssim {\rm [O/Fe]} \lesssim 0.26$. In the MW, thick-disc stars have a larger oxygen abundance than thin-disc stars with the same [Fe/H], as shown by, e.g. \citet{Franchini:2021aa}. For [Fe/H] $= -0.5$, the mean abundance ratios are [O/Fe] = $0.36 \pm 0.19$ and $0.24 \pm 0.07$ for the thick- and thin-disc stars, respectively \citep[see][]{Reddy:2006aa, Bertan:2016aa}. The mean [O/Fe] ratio for the possible NSCPs is $\sim$0.15, whereas the mean [O/Fe] for the thin-disc stars in the simulation is $\sim$0.13. Hence, from the [O/Fe] ratio in our sample, we can  deduce that most of the stars within the NSCPs, especially the ones with a low [O/Fe] ratio, could have originated out of a thin disc (see Section~\ref{sec:thin} for further reasoning).

All the NSCPs are still star-forming at $z = 4.4$, which could indicate that we are looking at a hybrid formation scenario of the MW's NSC if still star-forming clusters will inspiral in and bring the star-forming gas to the galactic centre. For all clusters at $z = 4.4$, the SF episodes are between 0.7 and 1.2~Gyr long, with a mean specific SF rate (SFR) between 0.8 and 1.2~Gyr$^{-1}$. The fact that the NSCPs are still star-forming again shows that the NSCPs are different from the proto-GCs discussed in \citetalias{Donkelaar:2022ab}, as we expect a burst of SF for GCs. At $z=4.4$, all clusters experience their highest SFR since birth, with SFRs between $10^{-3.9}$  and $10^{-1.6}$~M$_{\sun}$~yr~$^{-1}$.

\subsubsection{Thin-disc stars}\label{sec:thin}

\citet{Tamfal:2022aa} showed that multiple thin discs, defined as a stellar disc component with $v_{\phi} / \sigma_{\rm R}$ larger than unity, are already present at $z \sim 7$. These early thin discs will most likely evolve into the thick disc or bulge by $z=0$ (Van Donkelaar et al., in prep.). The detected stellar thin disc is rather compact in radial size at $z=4.4$, therefore the overall aspect ratio of the disc is larger than in present-day counterparts. However, this still means that these early thin discs could have had a big influence on the evolution of the galaxy. This is also shown by Figure~\ref{fig:prop}, as approximately half of the NSCPs have the fraction of ``thin disc-born'' stars higher than 0.5. The ``thin disc-born'' stars are defined by applying the \textsc{DBSCAN} \citep[see][]{Ester:1996aa} clustering algorithm to our simulation at different time steps.\footnote{See \citet{Tamfal:2022aa} for an explicit approach of identifying stars as thin-disc stars.} This way, sequences of mutually spatially connected particles are identified with a process comparable to that of a scatter kernel interpolation in SPH. The ID of the stars born in a thin disc is saved and from this we can calculate the fraction of ``thin disc-born'' stars, $F_{\rm thin}$. The NSCPs with $F_{\rm thin} > 0.5$, indicated with the black outline in Figure~\ref{fig:prop}, have a higher mean stellar metallicity ([Fe/H] $\sim -0.22$) and a lower [O/Fe] ratio ([O/Fe] $\sim 0.12$) than the mean of the whole sample, which is expected from thin-disc stars \citep[see, e.g.][]{Franchini:2021aa, Reddy:2006aa, Bertan:2016aa}. Combining the birth environment determined by \textsc{DBSCAN} and the discrepancy between the metallicity properties of the two populations of NSCPs in Figure~\ref{fig:prop}, we can confidently say that the  stars identified using \textsc{DBSCAN} are indeed ``thin disc-born'' stars. The NSCPs consist of a minimum of 3 per cent in mass out of ``thin disc-born'' stars. Furthermore,  for 47 per cent of the clusters (22), the ``thin disc-born'' stars make up more than half of their stellar mass at $z = 4.4$.

The different properties of these two types of NSCPs could suggest that the NSCPs with $F_{\rm thin} >0.5$ formed under different conditions. This is further investigated in Figure~\ref{fig:thindisc}, where the NSCPs are plotted in the $x$-$y$-$z$-space with the stellar galactic angular momentum pointing in the $+z$ direction at $z = 4.4$. The NSCPs form an elliptical ring around the galactic centre of the main galaxy halo. From the Figure, it is clear that the two populations of NSCPs, high- and low-$F_{\rm thin}$, can be found in different regions in this elliptical ring and there is barely any mixing between the two groups. The colour bar in Figure~\ref{fig:thindisc} shows the mean age of the stars within the NSCPs, from which we can deduce that the high-$F_{\rm thin}$ NSCPs have an older mean stellar age. We find a mean stellar age for high-$F_{\rm thin}$ NSCPs of $250.2 \pm 45.2$~Myr. For low-$F_{\rm thin}$ NSCPs, we find a mean stellar age of $102.1 \pm 45.1$~Myr. Notably, the ``thin disc-born'' stars are the oldest stars within the stellar clusters. The mean age of the NSCP stars is therefore correlated with where the cluster can be found within the elliptical ring at $z = 4.4$. Tantalizingly, there is no clear relation between the mean age of the ``thin disc-born'' stars within the NSCPs and the birth radius of the oldest star within the cluster, even though some clusters include stars born far outside the disc.

As expected, the SFHs of the two NSCP groups, high- and low-$F_{\rm thin}$,  are also different. The $F_{\rm thin} >0.5$ NSCPs start forming stars  earlier, consistent with the fact that these clusters are older. However, they also show an approximately constant SFR of $\sim$$0.4 \times 10^{-2}$~M$_{\sun}$~yr~$^{-1}$ from $z \lesssim 6$, with an increase in SFR around $z \sim 4.5$. On the contrary, the $F_{\rm thin} \leq 0.5$ NSCPs have on average a lower SFR and show this increase in SF from $z=5$.

\subsection{The hybrid formation scenario}\label{sec:hybrid}

As discussed in Section~\ref{sec:intro}, GC accretion and in-situ SF are both expected to contribute to the build up of NSCs over a Hubble time \citep[see, e.g.][]{Guillard:2016aa}. The NSC of the MW is a prominent example, as it shows both young stars likely formed in-situ and old, metal-poor components that possibly were accreted from GCs \citep[e.g.][]{Antonini:2012aa, Feldmeier:2015aa, Feldmeier:2020aa, Arca:2020aa}. We have shown so far that the process of infalling gas-rich stellar clusters will likely contribute to the old, metal-poor component of the NSC in our simulated galaxy. Nevertheless, we have also shown that subsequent in-situ SF will be needed to acquire a stellar metallicity that is consistent with that of the MW's NSC.

\begin{figure}
\centering
\setlength\tabcolsep{2pt}%
\includegraphics[ trim={0cm 0cm 0cm 0cm}, clip, width=0.48\textwidth, keepaspectratio]{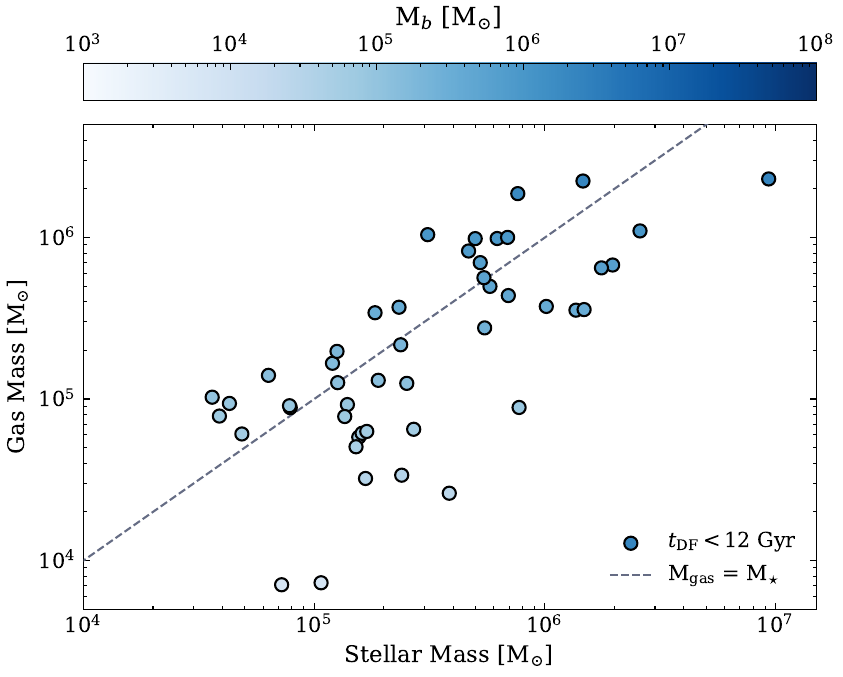}
\caption{Gas mass as a function of stellar mass. The 47 dots indicate the stellar cluster sample defined by applying the selection criterion described in Section~\ref{sec:class}  and additionally imposing $t_{\rm DF} < 12$~Gyr (Section~\ref{sec:dynamicaltime}). The colour bar represents the total baryonic mass, $M_{\rm b}$,  of the stellar clusters. The gray, dashed line displays the 1:1 ratio between the stellar and gas mass.}
\label{fig:mgas}
\end{figure}

For the in-situ SF formation scenario, gas will need to fall into the nucleus and then transform into stars \citep[e.g.][]{Loose:1982}. This gas reservoir could come from inflowing gas due to, e.g. non-axisymmetric galactic features \citep[e.g.][]{Milsavljevi:2004aa,Bekki:2007aa}, but could also come from the NSCPs themselves. From the stellar-gas mass relation shown in Figure~\ref{fig:mgas}, one can conclude that the clusters with a high baryonic mass still include a significant amount of gas. The NSCPs have a $F_{\rm \star}$ between $0.19$ and $0.94$, with a mean stellar mass fraction of $0.56$. One could assume that the gas from these clusters that is not yet converted into stars when the cluster reaches the galactic centre would become part of the gas reservoir used for the central in-situ SF.

We can obtain an estimate on the maximum amount of stellar mass that can assemble from in-situ SF from gas coming from infalling stellar clusters. We here assume that only the stellar clusters that have a DF time of $t_{\rm DF} \leq 1$~Gyr are still gas-rich when reaching the galactic centre of the galaxy and find a total gas mass that can be used for SF in the central region to be $10^{7.0}$~M$_{\sun}$. We thus find that in-situ SF of gas contributed from NSCPs contributes to a maximum 20 per cent of the total baryonic mass of the clusters with  $t_{\rm DF} < 12$~Gyr ($10^{7.7}$~M$_{\sun}$). However, as these stars will form very close to the time when the stellar clusters were accreted, this type of in-situ SF will only contribute old stars to the NSC.

\begin{figure}
\centering
\setlength\tabcolsep{2pt}%
\includegraphics[ trim={0cm 0cm 0cm 0cm}, clip, width=0.47\textwidth, keepaspectratio]{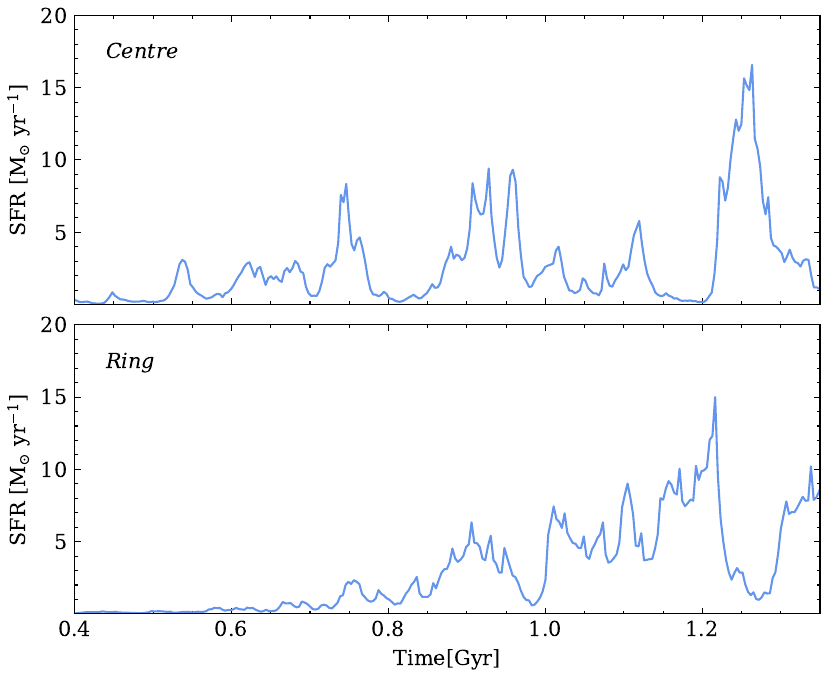}
\caption{The evolution of the SFR within the galactic centre and within the NSR. We defined the galactic centre as a sphere of 100~pc around the centre of mass of the main galaxy and the NSR to be the region between 200~pc and 1~kpc around the galactic centre of the simulated galaxy in the $y$-$z$-plane and with a height of 500~pc.}
\label{fig:SFR}
\end{figure}

Because the NSCs of massive galaxies show significant contributions from young and enriched populations \citep[e.g.][]{Kacharov:2018aa, Pinna:2021aa, Fahrion:2021aa, Fahrion:2022aa}, the galactic-scale non-axisymmetric structures such as bar and spiral structures could funnel towards the centre diffuse gas, providing new fuel for in-situ SF creating the younger stars detected within the MW's NSC. This extra channel is further supported by the evolution of the SFR within the galactic centre of the main galaxy, shown in the top panel of Figure~\ref{fig:SFR}, as SF is already happening within the central region of the galaxy. Furthermore, because high-mass NSCs have a high ellipticity \citep[e.g.][]{Seth:2006aa, Spengler:2017aa, Georgiev:2014aa}, one could infer that such ellipticity is more easily explained by in-situ SF from inflowing gas \citep[see, e.g.][]{Spengler:2017aa}. This is because elongation could originate from rotational flattening, which requires a dynamically important coherent angular momentum on the star-forming baryons, which is easier to maintain through dissipative accretion in the galactic nucleus than from  non-dissipative merger dynamics of pre-existing stellar clusters.

\begin{figure}
\centering
\setlength\tabcolsep{2pt}%
\includegraphics[ trim={0cm 0cm 0cm 0cm}, clip, width=0.495\textwidth, keepaspectratio]{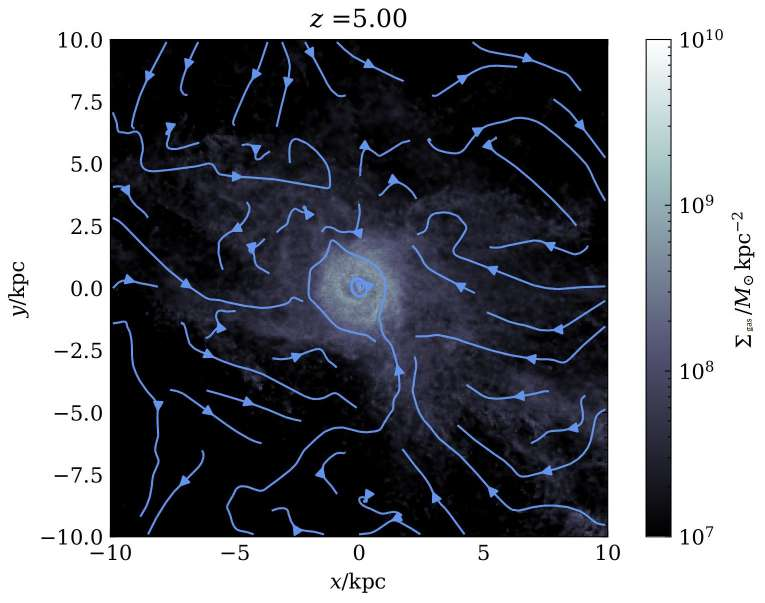}
\caption{Gas surface density map of the main galaxy at $z=5.0$, with the velocity vectors of the gas overlaid. The galaxy is centred face-on on the disc stars from the main galaxy halo.}
\label{fig:gasstreams}
\end{figure}

Furthermore, the gas that has been directed toward the centre could settle onto an elliptical ring \citep[see Section~\ref{sec:ring}; see, e.g.][]{Contopoulos:1989aa, Binney:1991aa,Knapen:1999aa, Regan:2003, Li:2017aa, Sormani:2022aa,  Rebecca:2022aa}. The simulation displays bar formation \citep[as discussed in][]{Tamfal:2022aa}: this could therefore be another mechanism that will contribute to the gas reservoir in the centre needed for in-situ SF. Figure~\ref{fig:gasstreams} displays the gas surface density\footnote{The depth of the surface density plots is the depth of the simulated box.} at $z=5.0$ with the velocity vectors of the gas overlaid.  We can identify a couple of gas streams coming from the outer parts of the galaxy towards the centre.

Concluding, we would expect to have three channels that can contribute to the total stellar mass of the NSC: (i) gas-rich stellar cluster accretion brings in stars formed outside of the galactic nucleus and (ii) contributes gas fuel to the central reservoir for subsequent in-situ SF (both these channels producing relatively old stars); additionally, (iii) galactic-scale non-axisymmetric structures such as  bar and spiral structures funnel towards the centre diffuse gas (not previously in clusters), which can then provide fuel for in-situ SF, producing both old and young stars (although this was not directly simulated).  Hence, we have shown that the in-situ SF component of the hybrid formation process is occurring through two distinct channels. Further research is needed to determine the relative importance of each channel, as the formed stars (both directly from clusters and in-situ SF) could also contribute to the build-up of other components of a galactic centre, e.g. a pseudo-bulge or NSR, instead of the NSC. Based on the total stellar mass and metallicity, it is plausible to hypothesize that not all of the stars within the NSCPs will ultimately become part of the NSC. It is likely that some of these stars will be integrated into other components of the galactic centre. Further research will be necessary to investigate the fate of these stars and to determine the proportion that will ultimately contribute to the NSC.

\subsection{The nuclear stellar ring}\label{sec:ring}

\begin{figure}
\centering
\setlength\tabcolsep{2pt}%
\includegraphics[ trim={0cm 0cm 0cm 0cm}, clip, width=0.48\textwidth, keepaspectratio]{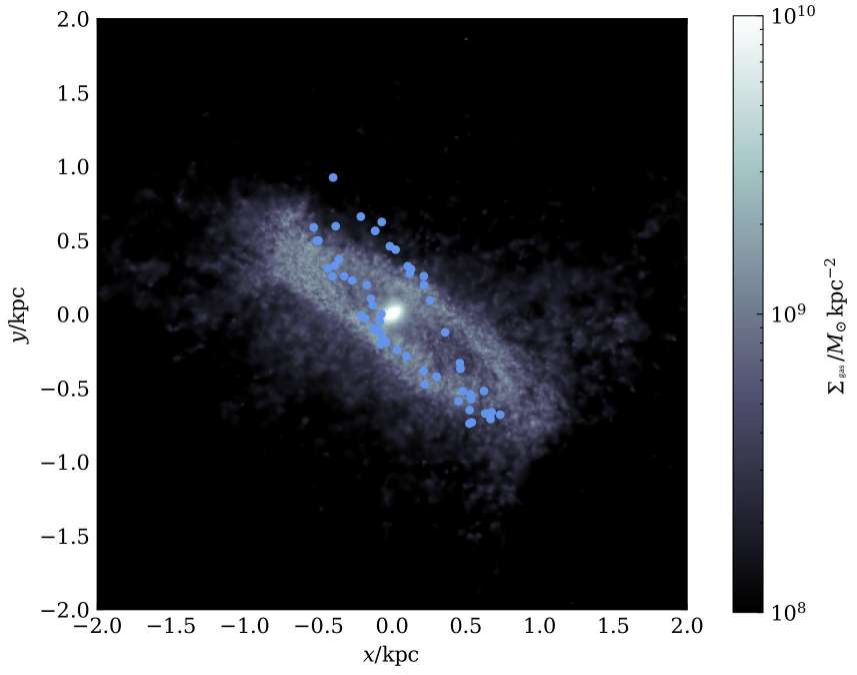}
\caption{Gas surface density map of the main galaxy at $z=4.4$. The dots represent all identified clusters within $r<1.5$~kpc. The stellar galactic angular momentum is pointing in the $+y$ direction.}
\label{fig:Nring}
\end{figure}

In addition to the  NSC, it is known that the MW has also an NSD-like structure \citep[e.g.][]{Schodel:2021aa, Schultheis:2021aa}. Figure~\ref{fig:Nring} projects the stellar clusters without the DF time-scale constraint found within 1.5~kpc from the galactic centre at $z = 4.4$ on the gas surface density of the galaxy. This shows that there is a striking resemblance between the location of the stellar clusters and what one would identify as an NSR. From Section \ref{sec:dynamicaltime}, we determined that most of the NSCPs will have fallen to the galactic centre  by $z = 0$. However, this estimation did not account for mutual interactions between the stellar clusters. Consequently, at $z=0$ there could still be a few of the stellar clusters in the spherical ring surrounding the NSC.

The ring visible in Figure~\ref{fig:Nring} has a larger radius and scale-height than those of the present-day NSD of the MW \citep[$R \approx 220$~pc and $h \approx 50$~pc;][]{Launhardt:2002aa, Nishiyama:2013aa, Nogueras:2020aa, Gallego:2020aa}. Nevertheless, the radius of the NSR could decrease over time due to DF strengthened by the bar-induced resonances \citep[see][]{Bortolas:2022aa}. Additionally, a ring forming at the inner ends of bar-torque-produced dust lanes, as shown in \citet{Kim:2012aa}, will shrink in size by 10 to 20 per cent as collisions of dense clumps inside the ring take away angular momentum from the ring. Therefore, we can assume that, at $z=0$, the ring shown in Figure~\ref{fig:Nring} will most likely have a radius and scale-height closer to the MW's NSD's values.

The MW's NSD local orbital inclination relative to the disc’s midplane is between $7^{\circ}$ and $14{^\circ}$, with a mean of $\sim$$10^{\circ}$ \citep[][]{Gillessen:2009aa}. The ring displayed in Figure~\ref{fig:Nring} is on a $\sim$$14^{\circ}$ inclination relative to the disc stars, which is thus in correspondence with the observations of the MW. Furthermore, both in the MW and the simulation, the stellar orbits are on average not circular in the nuclear ring and disc. The mean ellipticity of the MW's nuclear disc inferred by \citet{Bartko:2009aa} is 0.37 $\pm 0.07$, whereas the ellipticity of the ring formed by the NSCPs in the simulation is 0.23.

The gas surface density plot in Figure~\ref{fig:Nring} also depicts a gaseous ring surrounding the galactic centre of the main galaxy. This is expected from NSRs, since these rings are most likely produced by the radial infall of gas caused by angular momentum loss as a consequence of  their nonlinear interactions with an underlying stellar bar potential \citep[e.g.][]{Combes:1985aa, Athanassoula:1992aa, Buta:1996aa, Patsis:2010aa, Kim:2012ab}. Tantalizingly, the orbits of the stellar clusters are not aligned with this gaseous ring, which has an inclination relative to the disc stars of  $\sim$$14^{\circ}$.  Numerical simulations of SF in the nuclear ring find that the stellar clusters and gaseous ring overlap, and have a similar inclination relative to the disc's midplane \citep[e.g.][]{Seo:2013aa, Seo:2014aa, Moon:2021aa}. This difference can be explained by the different formation environments of the stellar clusters.  The stellar clusters within the simulations of \citet{Seo:2013aa, Seo:2014aa} were entirely formed within the ring, whereas we show that most of the NSCPs in this simulation include stars that have migrated from the stellar disc towards the nuclear region. Therefore, the inclination of the clusters relative to the gaseous ring could be due to the different environmental effects both systems have experienced and differing initial angular momentum.

The total gas mass of the NSR is $3.4 \times 10^8$~M$_{\sun}$, which is in line with what is observed in NSRs \citep[$\sim$$1$ to $6 \times 10^8$~M$_{\sun}$;][]{Buta:2000aa, Benedict:2002aa, Sheth:2005aa, Schinnerer:2006aa}. The bottom panel of Figure~\ref{fig:SFR} plots the temporal evolution of the SFR in the ring. As expected from the literature, the SFR of the NSR is within a range of $\sim$1 to $20$~M$_{\sun}$~yr$^{-1}$ \citep[][]{Mazz:2008aa, Comeron:2010aa}. We find an SFR in the nuclear ring that is slowly increasing with time, with short bursts of SF with time intervals between the bursts of roughly $\sim$50~Myr.

\section{Discussion}\label{sec:disc}

We explored the possibility of stellar clusters within 1.5~kpc of the main galaxy halo at $z>4$ to contribute to the old, metal-poor
component of the NSC, and their link to the thin disc and the NSR, through the use of the $N$-body, hydrodynamical, cosmological ``zoom-in'' simulation GigaEris \citep[][]{Tamfal:2022aa}. There is no guarantee that all the stellar clusters identified as NSCPs in this paper will become part of the NSC at present day, as the calculation in Section~\ref{sec:dynamicaltime} is approximate. This, for example, does not take into account the interaction between the stellar clusters or the influence of galactic structures like the bar and spirals, meaning that in reality the time frame for a cluster to reach the galactic centre could be different. Additionally, as discussed in Section~\ref{sec:hybrid}, these stellar clusters could contribute to other components of the nuclear region, e.g. the pseudo-bulge. Nevertheless, as the goal of this paper is to show the possibility of the MW's NSC being formed using the hybrid scenario and the link with the thin-disc stars, making these assumptions for the DF time-scale is reasonable.

Although we are unable to follow the evolution of the selected stellar clusters to $z=0$ and, therefore, cannot make definitive statements on whether these clusters are truly NSCPs and thus contribute to the formation of the old, metal-poor component of the NSC, many of the discussed properties are grossly consistent with the present-day NSC of the MW. One clear difference from observations is the metallicity of the stars in the NSCPs in our simulation, which on average is slightly lower than the average metallicity of the stars in the MW's NSC \citep[e.g.][]{Schodel:2020aa}. This reinforces the likelihood that these stars will contribute to the older, metal-poor component. Additionally, as we argue for a hybrid formation scenario, the in-situ SF from the gas funneled from the galactic-scale non-axisymmetric structures -- the third channel discussed in Section~\ref{sec:hybrid} -- will increase the average metallicity of the stars in this region.  In the case of late-type host galaxies such as our MW, both the GC accretion and in-situ SF scenario are expected to work in parallel, as supported by the MW’s NSC’s observed metallicity spread and complex SFH \citep[e.g.][]{Do:2015aa}. However, the contribution of each mechanism to the build-up process of the MW's NSC is still not clear.

The cumulative stellar mass at redshift $z=4.4$ attributed to the NSCPs in our simulation amounts to $10^{7.5}$~M$_{\sun}$. To compare this number to current observations, one can use the mixed-metallicity model for the SFH of the MW's NSC described by \citet{Schodel:2020aa}. This model shows that only $28$ per cent of the total current stellar mass formed in the past 11~Gyr (i.e. since $z \sim 2.5$). Under the assumption of a constant SFR within the clusters in the $\sim$1~Gyr between $z=4.4$ and $z = 2.5$, and taking the highest value of the SFR observed in those clusters at $z=4.4$ ($10^{-1.6}$~M$_{\sun}$~yr~$^{-1}$), the combined stellar mass of the NSCPs would reach $10^{7.8}$~M$_{\sun}$ by $z = 2.5$ and $10^{7.9}$~M$_{\sun}$ by $z = 0$. However, it is noteworthy that the total baryonic mass of the NSCPs is only $10^{7.7}$~M$_{\sun}$, indicating that the maximum stellar mass contribution from the NSCPs could only be $10^{7.7}$~M$_{\sun}$ (in other words, the quoted SFR cannot be sustained for 1~Gyr), yielding a $z = 0$ value of $10^{7.8}$~M$_{\sun}$. This latter number exceeds the presently estimated stellar mass of the MW's NSC, which is currently reported to be a few $10^7$~M$_{\sun}$ \citep[e.g.][]{Launhardt:2002aa, Feldmeier:2014aa, schodel:2014ab, Fritz:2016aa, Fledmeier:2017aa}.

These numbers can only be considered a rough estimate, as our understanding is limited beyond the final snapshot of our simulation at $z=4.4$. As outlined in Section~\ref{sec:hybrid}, our discussion emphasizes the probable existence of a third formation channel involving inflowing gas. Therefore, we suggest that the baryonic mass contribution from NSCPs (i.e. the first two channels) constitutes only a portion of the total stellar mass of the NSC at $z \sim 0$. However, determining this specific fraction remains uncertain and will require further research.

Hence, it is likely that (i) only a fraction of the NSCPs will reach the galactic centre, and, for the same mass considerations, (ii) only a portion of the SF in the centre will contribute to the total mass of the NSC (some formed stars could potentially become part of the pseudo-bulge, as argued by \citealt{Tamfal:2022aa}). Furthermore, studies indicate that the MW's NSC has undergone stellar mass loss due to stellar winds \citep[e.g.][]{Blum:2003aa, Pfuhl:2011aa, Das:2021aa}. For instance, the work by \citet[figure~19]{Blum:2003aa} showed that stars formed between 5 and 12~Gyr ago experienced a loss of approximately 68~per cent. Furthermore, the GigaEris simulation's main halo has been designed to have a halo virial mass of $1.4 \times 10^{12}$~M$_{\sun}$ at $z=0$, aligning with the upper end of the estimated mass of the MW \citep[see, e.g.][]{Wang:2020aa, Bobylev:2023aa, Jiao:2023aa}. Since the NSC's mass is expected to correlate well with the galaxy's mass \citep[see, e.g.][]{Fahrion:2022aa}, a more massive NSC of $M_{\rm NSC} \gtrsim 10^8$~M$_{\sun}$ can be anticipated \citep[e.g.][]{Iskren:2016aa, Fahrion:2022aa}, supporting the feasibility of the hybrid theory presented in this paper.

\subsection{The survival of gas}

In the NSC formation theory discussed in this paper, we assume that the infalling gas-rich NSCPs add to the gas reservoir in the centre of the galaxy that is needed for in-situ SF, the first channel in Section~\ref{sec:hybrid}. For this scenario to work, the gas of the NSCPs should still be part of the clusters at the time of falling in and still be star forming. The first requirement can be investigated using the tidal radius of the clusters. The tidal radius is the distance from a satellite orbiting in a host potential beyond which its material is stripped by the tidal force. All the NSCPs are outside of the tidal radius at $z = 4.4$.

As discussed in Section~\ref{sec:method}, each star particle is created stochastically with an initial mass of $m_{\star} = 1026$~M$_{\sun}$ using a simple gas density of $n_{\rm SF} > 100$~$m_{\rm H}$~cm$^{\text{-} 3}$ and temperature threshold criterion of $T_{\rm SF} < 3 \times 10^4$~K. We find that, on average, 48 per cent of the gas within the NSCPs is below this temperature and above the density threshold. Therefore, there is a reasonable chance that the gas-rich stellar clusters can contribute gas fuel to the central reservoir for subsequent in-situ SF.

\subsection{Radial migration of stars}

Most of the stars within the NSCPs were born at around $\sim$1~kpc from the galactic centre. Nevertheless, there is a small group of stars that were born further away. Before all the bound stars of the NSCPs become part of the elliptical ring around the galactic centre, as shown in Figure~\ref{fig:thindisc} at $z =4.4$, the stars born far outside the galactic centre within these clusters must have migrated towards the centre of the galaxy.

We followed the stellar particles for a couple of NSCPs from $z = 6.91$ to $z=4.4$. We find that, on average, roughly $\sim$20~per cent of the stars were born outside the cluster and migrated ``alone''\footnote{In the code, a stellar particle represents an entire stellar population, meaning that when one stellar particle migrates, it is in reality an ensemble of stars. The stars are thus never really alone.} towards the cluster. These stellar particles often formed in the thin disc. For most stars, the migration towards the centre occurs around $z\sim 6.8$. In the galactic centre they will become part of the main NSCPs and  correspond on average to $\sim$0.05~per cent of the total final mass of the NSCPs. The exact dynamics and kinematics of the radial migration at $z>4$ will be studied in a subsequent paper (Van Donkelaar et al., in prep.).

\subsection{Comparison to literature}

While our work is one of the first to discuss the formation of an NSC using a high-resolution, cosmological ``zoom-in'' simulation, our results are mostly consistent with those from previous numerical simulations \citep[e.g.][]{Antonini:2013aa, Seo:2013aa, Seo:2014aa, Gnedin:2014aa, Li:2015aa, Guillard:2016aa, Tsatsi:2017aa, Brown:2018aa, Seo:2019aa, Leaman:2022aa, Sormani:2022aa}. For example, \citet{Gnedin:2014aa} found that for the MW the most probable value for the mass of the central cluster formed by the infall of GCs is 2--$6 \times 10^7$~M$_{\sun}$, which is similar to the combined stellar mass of our NSCPs, $3.2 \times 10^7$~M$_{\sun}$. \citet{Brown:2018aa} show that the observed abundance in NSCs must involve additional sources next to the infall of GCs, which is similar to our finding that currently the metallicity of the NSCPs is too low and will need additional channels, e.g. our third channel, to increase.

Furthermore, in our simulation we find an SFR in the nuclear ring that is slowly increasing with time, with short bursts of SF between $\sim$1 and $20$~M$_{\sun}$~yr$^{-1}$. The time intervals between the bursts are roughly $\sim$50~Myr. This is consistent with \citet{Seo:2013aa}, who found that the SFR in nuclear rings displays a single primary burst followed by a few secondary bursts with a time interval between the bursts to be roughly $\sim$50--80~Myr. Interestingly, both \citet{Seo:2013aa} and \citet{Seo:2019aa} found that the SFR will decrease during the simulation, whereas we see an increase in SFR over time. In \citet{Seo:2013aa}, the SFR decreases to small values after the second burst. This difference is probably due to the fact that they start with an infinitesimally thin, rotating disc, which is unmagnetized and isothermal. We, on the other hand, used a high-resolution, cosmological ``zoom-in'' simulation of an MW-sized galaxy to investigate the NSR. It could be that \citet{Seo:2013aa} showed what would happen after $z=4.4$, as our simulation ends with a thin-disc-like structure.

The nuclear ring detected in the simulation has a larger radius than that of the present-day NSD of the MW, but also larger than the simulations performed by for example \citeauthor{Seo:2019aa} (\citeyear{Seo:2019aa}; less than $\sim$0.6~kpc). Nevertheless, \citet{Li:2017aa} used hydrodynamic simulations with static stellar potentials to show that a nuclear ring forms only in models with a central object exceeding $\sim$1 per cent of the total disc mass, and that the ring size increases almost linearly with the mass of the central object. As the simulation already has a dense central region, from which most of the stars will become part of the pseudoubulge, it opens the possibility that the presence of a massive compact star-forming region or pseudobulge would make a ring large when it first forms. The formed NSR can even become larger as it grows due to an addition of gas with larger angular momentum from outer regions, as discussed in \citet{Seo:2019aa}.

\section{Conclusions}\label{sec:conc}

Using a high-resolution, cosmological ``zoom-in'' simulation of an MW-sized galaxy halo \citep[GigaEris;][]{Tamfal:2022aa}, we have investigated the possibility of stellar clusters within 1.5~kpc of the main galaxy halo at $z>4$ to contribute to the old, metal-poor component of the NSC, and their connection to the thin disc. Moreover, we explored the relationship between the NSC and the NSD. Our main conclusions are as follows:

\begin{itemize}

    \item We define NSCPs as clusters within 1.5 kpc from the centre of the main galaxy and a $t_{\rm DF} <12$~Gyr. The total stellar mass of the clusters together, $\sim$$10^{7.5}$~M$_{\sun}$, is in the same order of magnitude as the observed mass of the MW's NSC.
    
    \item NSCPs at $z=4.4$ have a relatively low stellar metallicity, $-0.47 \lesssim {\rm [Fe/H]} \lesssim -0.11$, in comparison to the stars in the MW's NSC. This reinforces the likelihood that these stars will contribute to the older, metal-poor component. Additonally, the stellar metallicity of the MW's NSC can be increased by the in-situ SF later during the formation. The detected stellar metallicity  makes the selected star clusters in this paper very different from the proto-GCs discussed in \citetalias{Donkelaar:2022ab}.
    
    \item The NSCPs have a low oxygen-to-iron ratio,  $0.02 \lesssim {\rm [O/Fe]} \lesssim 0.26$, which points to the fact that a fraction of the stars within the clusters are born within the thin disc \citep[e.g][]{Reddy:2006aa, Bertan:2016aa,Franchini:2021aa}. Combining the birth environment of the stars with the oxygen-to-iron ratio, we have determined that the NSCPs consist of a minimum of 3 per cent out of ``thin disc-born'' stars, and for 43 per cent of the clusters the ``thin disc-born'' stars make up more than half of the total stellar mass at $z = 4.4$.
    
    \item The NSCPs form an elliptical ring around the galactic centre of the main galaxy halo at $z = 4.4$. The NSCPs with $F_{\rm thin} > 0.5$ can be found in different regions than the low-$F_{\rm thin}$ clusters in this ring and have a higher iron-to-hydrogen ratio. Furthermore, the NSCPs with $F_{\rm thin} > 0.5$ have a higher mean age of stars in the cluster than the other clusters.
    
    \item The hybrid formation scenario for the NSC of an MW-sized galaxy is the most likely. This is the result of three channels that contribute to the total stellar mass of the NSC. Gas-rich stellar cluster accretion brings in stars formed outside of the NSC and adds to the gas reservoir in the centre needed for in-situ SF. Conjointly, the galactic structures like the disc and the bar will funnel the gas towards to centre, which can also be used for in-situ SF.
    
    \item  Assuming that all the gas from the high-mass infalling stellar clusters with a $t_{\rm DF} \leq 1$~Gyr will turn into stars within the NSCPs, we find that in-situ SF contributes to a maximum of 30 per cent  of the stellar mass of the NSC cluster, only including the baryonic mass added from the NSCPs, of an MW-sized galaxy.
    
    \item We detect an NSR within the GigaEris simulation with a radius of $\sim$500~pc at $z = 4.4$. This ring hosts the NSCPs at this redshift and has periodic SF short bursts of $\sim$1 to $20$~M$_{\sun}$~yr$^{-1}$, with time intervals of $\sim$50~Myr.
    
\end{itemize}

\section*{Acknowledgements}
We thank the anonymous Reviewer for their constructive input. PRC, LM, and FvD acknowledge support from the Swiss National Science Foundation under the Grant 200020\_207406. The simulations were performed on the Piz Daint supercomputer of the Swiss National Supercomputing Centre (CSCS) under the project id s1014. FvD would like to thank Katja Fahrion for initiating the idea that the clusters in the centre could be part of the NSC and useful discussions along the way. 

\section*{Data Availability}
The data underlying this article will be shared on reasonable request to the corresponding author.

\bibliographystyle{mnras}
\bibliography{paper} 

\bsp	
\label{lastpage}
\end{document}